\documentclass[preprint,12pt,3p]{elsarticle}

\usepackage{amssymb}

\journal{New Astronomy Reviews}

\begin{document}

\begin{frontmatter}

\title{Kepler-9: the First Multi-Transiting System and the First Transit Timing Variations}

\author[label1]{Darin Ragozzine}
\address[label1]{Brigham Young University, Department of Physics and Astronomy, N283 ESC, Provo, UT 84602, USA}

\ead{darin\_ragozzine@byu.edu}

\author[label2]{Matthew J. Holman}
\address[label2]{Harvard-Smithsonian Center for Astrophysics, 60 Garden St., MS 51, Cambridge, MA 02138, USA}

\begin{abstract}
Kepler-9, discovered by Holman et al. 2010, was the first system with multiple confirmed transiting planets and the first system to clearly show long-anticipated transit timing variations (TTVs). It was the first major novel exoplanet discovery of the Kepler Space Telescope mission. The Kepler pipeline identified two Saturn-radius candidates (called Kepler Objects of Interest or KOIs): KOI-377.01 with a 19-day period and KOI-377.02 with a 39-day period. Even with only 9 transits for KOI-377.01 and 6 of KOI-377.02, the transit times were completely inconsistent with a linear ephemeris and showed strongly anti-correlated variations in transit times. Holman et al. 2010 were able to readily show that these objects were planetary mass, confirming them as \emph{bona fide} planets Kepler-9b and Kepler-9c. As a multi-transiting system exhibiting strong TTVs, the relative planetary properties (e.g., mass ratio, radius ratio) were strongly constrained, opening a new chapter in comparative planetology. KOI-377.03, a small planet with a 1.5-day period, was not initially discovered by the Kepler pipeline, but was identified during the analysis of the other planets and was later confirmed as Kepler-9d through the BLENDER technique by Torres et al. 2011. Holman et al. 2010 included significant dynamical analysis to characterize Kepler-9's particular TTVs: planets near resonance show large amplitude anti-correlated TTVs with a period corresponding to the rotation of the line of conjunctions and an additional "chopping" signal due to the changing positions of the planets. We review the historical circumstances behind the discovery and characterization of these planets and the publication of Holman et al. 2010. We also review the updated properties of this system and propose ideas for future investigations.

\end{abstract}

\begin{keyword}
exoplanets \sep history \sep Kepler Space Telescope
\end{keyword}

\end{frontmatter}



\section{Introduction}
\label{intro}
Shortly after the first confirmed transiting planet HD209458b in 2000 \citep{2000ApJ...529L..45C}, the field quickly mobilized to identify a wide variety of scientific investigations enabled by such planets. \citet{2000ApJ...529L..45C} itself notes the possibility of additional transiting planets in the system, though with 20/20 hindsight we now know that additional transiting planets in Hot Jupiter systems are very rare (e.g., \citep{2012PNAS..109.7982S}). These investigations were often extensions and updates of techniques that had been used on eclipsing binaries for decades; for example, \citep{2003ASPC..294..449C} and \citep{2003sf2a.conf..149S} include the idea of ``transit timing'' presumably based on ``eclipse timing'' that was a well-established technique (see also the prescient \citep{1952Obs....72..199S}). \citet{2002ApJ...564.1019M} had already proposed that additional inclined planets could be detected by the orbital precession which would change the durations and times of transits while  \citep{2003ApJ...592..555B} noted that additional planets orbiting HD209458 would alter the transit times of the known planet.

In independent and simultaneous analyses, \citet{2005MNRAS.359..567A} and \citet{2005Sci...307.1288H} showed that deviations of the transit times were much more sensitive than deviations on the transit duration (which is much harder to characterize) and that even ground-based observations had sufficient precision to detect Earth-size planets in certain resonant configurations. Using the known transiting planet as a lever, Transit Timing Variations (TTVs) could detect planets smaller than any other technique.  \citet{2005Sci...307.1288H} also anticipated the possibility of ``double transiting'' systems and the opportunity that such systems would offer for determining the planets' physical and orbital properties.

Fueled with a new detection technique, many teams sought to obtain multiple transit timing measurements on the ever-growing list of exoplanets around bright stars. Due to the strong bias towards detecting Hot Jupiters and the challenges of ground-based observation of smaller planets, all the early searches for TTVs were on Hot Jupiter systems. No detections were made, but some interesting upper limits (e.g., excluding Earth-mass planets in the 2:1 resonance) were established.\footnote{Most Hot Jupiters are now known to not show clear TTVs because they have no nearby companions to cause TTVs \citep{2012PNAS..109.7982S}.} 

After years of ground-based attempts, every expectation was that the exquisite precision, prolonged duty cycle, and continuous cadence of observations by the Kepler Space Telescope (hereafter \emph{Kepler}) would enable the first detections of TTVs. As in many areas, \emph{Kepler} was no disappointment and its first major novel discovery was the detection of TTVs of the planets orbiting the star Kepler-9, originally ``Kepler Object of Interest'' KOI-377. This article discusses the historical aspects of the discovery, characterization, and publication of the Kepler-9 system in Holman et al. 2010 \citep{2010Sci...330...51H}, hereafter H+10. Both authors of this paper (DR and MH) were heavily involved in the original publication and used their notes (mostly in the form of emails) to describe what happened behind the scenes of this major discovery.

\section{Background}
\label{background}
The Kepler Space Telescope (hereafter \emph{Kepler}), designed to study the population of transiting exoplanets, was proposed well before such planets were first discovered. It was also proposed before we learned that there are many detectably large planets on relatively short orbital periods and often in multi-planet systems.\footnote{Early optimistic estimates of the exoplanet yield from \emph{Kepler} assumed that every star had a Venus and Earth and that \emph{Kepler} should therefore discover several hundred exoplanets; DR remembers scoffing at such an ``obvious'' overestimate.} By the time \emph{Kepler} launched, there were about 40 known multi-planet systems, all from Radial Velocity (RV) observations. Most of these systems were of gas giant planets with periods of hundreds to thousands of days and, thus, unlikely to be detectable as ``multi-transiting systems'' (MTSs) even by \emph{Kepler}. 

Some of these multi-planet systems were composed of smaller planets (similar to ice giants) on relatively short periods and a correlation between small planets and multiplicity was known (e.g., \citep{2010A&A...512A..48L}). These were called Compact Multiples or Super-Earth and Neptune (SEN) systems in a few publications.\footnote{DR proposed the moniker Systems with Tightly-spaced Inner Planets (STIPs) in 2012 which has gained some traction in the literature}. Since only a handful of small transiting planets were known, the lack of multi-transiting systems before \emph{Kepler} was not surprising.

In sum, systems where precise transit times could be measured (Hot Jupiters) showed no evidence for TTVs, while systems where TTVs could be expected were not known to transit, nor were ground-based observations sufficiently precise to characterize them. This frustrating standoff reached a possible breaking point with the announcement of HAT-P-13b and c, the first system with a transiting planet where a second planet was known from RVs \citep{2009ApJ...707..446B}. Dan Fabrycky, who would later be second author on the Kepler-9 paper, discussed explicitly the value of transit data on multiplanet systems around this same time \citep{2009IAUS..253..173F}. 

DR started studying the HAT-P-13 system in detail and also recognized the challenge and value of MTSs. Knowing that early \emph{Kepler} results were available (though tightly-guarded) and hoping for some validation, he went to visit with Dave Latham who freely shared that \emph{Kepler} had already clearly discovered a number of systems with multiple candidate transiting planets. With only a little inside knowledge, DR went on to write Ragozzine \& Holman 2010 that thoroughly discussed the value of systems of multiple transiting planets \citep{RH10}.\footnote{Despite receiving a relatively favorable referee report, DR never resubmitted this paper as he felt too distracted analyzing actual \emph{Kepler} MTSs.} 

In particular, although it was known even by \citep{2005Sci...307.1288H}, \citep{RH10} discussed how scientifically valuable it was when the perturbing planet that caused TTVs was also transiting. When the perturber is not transiting, there are often significant degeneracies in the inference of the perturbing planet; furthermore, TTVs are almost independent of the mass of the perturbed transiting planet. As a result, TTVs in systems with multiple transiting planets are often the easiest way to measure the mass and radius of the same planet, yielding a density that provides crucial constraints on composition, formation, and habitability. 

At the same time that DR was writing \citet{RH10}, he was invited to formally be involved in the Kepler Science Team under the supervision of MH (his postdoctoral advisor). Around this time, the Science Team formed the Kepler TTV/Multis Working Group (KTMWG); this was originally going to be two separate groups, but there was so much overlap in interest that only one group really formed (see Steffen \& Lissauer, this volume). Jack Lissauer and Jason Steffen were chosen to be the first co-chairs and DR was chosen to be the scribe. At this time, all Kepler data and projects were still very carefully guarded, with practically no public knowledge, so those in the working group (and the broader science team) had a significant headstart and advantage on knowing what was coming. This was standard practice at the time as a way of validating results and supporting those who had spent significant amounts of time preparing, proposing, and operating the mission.

\section{Publication of Holman et al. 2010}

The Kepler pipeline was initially not good at finding additional transiting planets; many were identified by hand by Jason Rowe. His active searching and modeling of large numbers of planetary systems enabled a variety of early scientific invesigations in the KTMWG. 

In the KTMWG telecon on May 6, 2010\footnote{Meeting minutes indicate that the following participated in this telecon: Jack Lissauer, DR, Eric Ford, Jason Rowe, Bill Welsh, Jason Steffen, Dimitar Sasselov, Dave Latham, and Natalie Batalha}, Jason Rowe had just calculated transit times for \emph{Kepler} Quarters 0-3 and sent them to the group. He identified KOI-377 as the best system at this point showing TTVs. The two candidate planets, KOI-377.01 and KOI-377.02, had (best-fit) periods of 19.25 and 38.34 days, respectively. As a result, this first look at the data had only 7-8 transits of the inner planet and 4 transits of the outer planet, but the $\sim$2-minute precision of the \emph{Kepler} TTVs was far exceeded by the $\sim$20-40 minute signal. Based on the assumed stellar properties, these were relatively large planets (0.5 Jupiter radii). Jason Rowe saw that these planets had anti-correlated TTVs which were reasonable given the period ratio near 2:1 resonance. The star KOI-377 was noted to be relatively bright from a \emph{Kepler} perspective at 13.8 Kepler magnitude, but still quite faint for RV work. Other similarly strong TTV systems showed some evidence for being stellar in nature (e.g., \citep{2011MNRAS.417L..31S}), making KOI-377 the best planetary TTV candidate. 

Shortly after this telecon, MH took the lead on KOI-377 because studying TTV systems was the basis of his Kepler Participating Scientist Program. Dan Fabrycky (hereafter DF) and DR had worked on TTVs of the satellites of the the dwarf planet Haumea (2003 EL61) in the outer solar system and that experience would now be translated to KOI-377.  

DF, MH, and DR met on May 19 to discuss DF's initial fit to the TTV data. DF noted that, once removing a parabola, the TTVs show a $\sim$4-minute amplitude ``chopping'' which is caused by the alternating position of the outer planet at the time of transit of the inner planet. Due to the near 2:1 resonance, the Kepler-9 chopping signal manifested as offsets in every other transit of KOI-377.01. DF identified this as key for measuring the mass ratio of KOI-377.02 over the mass of the star, since this is essentially a detection of the barycentric motion of the star due to the changing position of the planet. Eric Ford emailed the KTMWG the same day with his own analysis of all the transit times from Jason and it quickly became clear that KOI-377 was the best case for TTVs that was clearly planetary in nature. 

There was a race to produce results by June since some early results were to be released at that time, based on the original data release plan. NASA allowed the mission to sequester 400 targets beyond June, but this was not large enough to hold back all planetary candidates, so there was discussion as to whether to release KOI-377 or keep it sequestered. The team decided to release 5 systems with multiple exoplanet candidates without confirming them as \emph{bona fide} exoplanets but presaging this new type of discovery (see \citep{2010ApJ...725.1226S} and Steffen \& Lissauer, this volume). We seriously considered trying to pull together a KOI-377 paper in two weeks, but eventually decided it would be better to keep this target sequestered. 

One challenge to assembling a paper rapidly was that the stellar properties of KOI-377 were not well constrained (as was the case in general for targets in the early Kepler days before much follow-up). This allowed the sizes of the planets to range from 0.5-1 Jupiter radii. Still, based on the clear dynamical interactions and dynamical stability, it was extremely likely that KOI-377.01 and KOI-377.02 were real planets. 

On May 20, we held a KTMWG telecon to discuss sequestering targets and/or coming up with results by the June data release date. In advance of the meeting, DF sent out some detailed notes with interesting updates on the KOI-377 system. 1) A quadratic fit to the TTVs showed that the best-fit period ratio was 2.023. At this point, we didn't know that just-wide-of-resonance period ratios are common\citep{2011ApJS..197....8L} and so DF assumed that this was an overestimate because of resonance libration. 2) Preliminary fits to the TTV data and to simulated systems showed that the observed anti-correlated quadratic TTVs are expected. 3) The ratio of the quadratic coefficients was well measured and directly related to the ratio of planet masses, by conservation of energy. This meant that the mass ratio would be measured quite well with the data in hand; based on simulated systems the ratio of planet masses was approximately 0.6 $\pm$ 0.1. (This remains true for all future models of these planets.) 4) Chopping was visible in KOI-377.01 and this would determine the mass of KOI-377.02. Preliminary fits gave a mass of 38 Earth masses for KOI-377.02. A planet with this low of a mass at the expected radius range had never been detected, but could be found on theoretical mass-radius diagrams. DF didn't know exactly why chopping was useful in constraining the mass because he had never considered a TTV analysis with two transiting planets before! (We were familiar with Nesvorny's TTV work \citep[e.g.,][]{2008ApJ...688..636N} which did investigate chopping-like TTV signals.) Despite these very nice results, there was still a little concern that we should check whether TTVs of this size could be caused by starspots. 

In this meeting, we set a goal for submitting a KOI-377 paper by June 15 and decide to proceed with the analysis. MH, DF, and DR were to work on determining planetary properties from the TTVs. DF was assigned to do dynamical stability to get maximum eccentricities. MH and Eric Ford were going to analyze the light curve to confirm Jason Rowe's transit times. Bill Welsh planned to work on astrophysical TTV systematics (e.g., starspots). We started planning the follow-up observations that we would need including, immediately, a spectrum for stellar classification, and eventually some basic RV data, not to confirm the planets, but as a consistency check. While we were optimistic about publishing in June, we decided to put KOI-377 on the sequester list because it was so awesome and we didn't want to take any chances about getting scooped. 

On May 23, Eric Ford reported on his own transit time fitting analysis that confirmed the key results from Jason Rowe. We got a very good look at Q1 data for the first time because that data had still not been corrected for light-travel-time (e.g., the Kepler-spacecraft time to Barycentric Julian Date), but Eric was able to use a conversion code from DR to get consistent times.  

On May 24, DR sent to MH and DF his first photodynamical model of the system. As far as we are aware, this is the first photodynamical model of a real exoplanetary system. DR confirmed that DF's best fit TTV models also worked well in the photodynamical model. 

DR stayed up late into the night to search the residuals of his photodynamical model for any additional planets. Using a Box Least Squares \citep{2002A&A...391..369K} code kindly provided by DF, DR identified a very strong signal at 1.5 days and the folded light curve looked very good. DR emailed the group and proposed that the new candidate be called KOI-377.03.\footnote{The next morning DR's wife Sarah saw prominently on his notes "I discovered a planet!". It remains DR's only planet discovery.}  

Within 24 hours, KOI-377.03 led to analysis and discussion by various members of the KTMWG. We found that KOI-377.03 looked good as a candidate, though it could still be a false positive. KOI-377.03 was noted to be too far away dynamically (period ratio over 10) from the other planets to make any difference in their TTVs (confirmed explicitly by later analyses). Similarly, the other planets would not induce significant TTVs in KOI-377.03, even with 10 years of \emph{Kepler} data. Jason Rowe explains why .03 was missed in the Kepler pipeline: ``The reason that KOI377.03 was not found is that the large TTVs produced large residuals when the transits of KOI377.01 and KOI377.02 are removed.  This tripped up the transit finding algorithm when searching for other candidates.'' The team resolves to search other TTV systems for additional planets.\footnote{In retrospect, using a photodynamical model to cleanly remove KOI-377.01 and .02 was not essential for BLS to discover .03, although correctly removing the transits would have made a difference for the Kepler pipeline. Interestingly, we note that strong TTVs have continued to affect the Kepler results: even in the final DR25 data release in 2018 \citep{2018ApJS..235...38T}, the vetter would reject the misfolded TTV data as not planetary in shape, leading to a ``false positive'' disposition for KOI-377.01. KOI-377.03 is also missed for basically the same reason as it was initially, though it was found by some intermediate versions of the Kepler pipeline. Overall, though non-Kepler pipeline searches have found a few additional planets in multi-planet systems, as far as the authors know, there have been no published systematic searches for planets using algorithms that do not assume a linear ephemeris \citep[such as QATS in][]{2013ApJ...765..132C}. DR participated in an unpublished search of more planets in KOIs using QATS lead by Ethan Kruse; his Florida Institute of Technology SPS1020 class identified a handful of promising candidates in 2015.} 

Since the \emph{Kepler} data were already months old before they got to us, we realized that a ground-based observation of KOI-377 could be very beneficial (but this never materializes). 

On May 25, we get our first ``reconnaissance'' spectrum from MacDonald Observatory, analyzed by Sam Quinn. The spectrum was pretty consistent with our expectations and with a solar-type star. 

On May 26, DF sent out his next step in the detailed analysis. Having obtained a best-fit coplanar case, he extended his analysis to allow the eccentricities to vary which led to a huge range of possible masses. We recognize this now as the mass-eccentricity degeneracy explored by \citep{2012ApJ...761..122L}, though Dimitri Veras was familiar with this result. At the time, it was disappointing because, although we could show the masses were in the planetary range, thus allowing TTVs to confirm these candidates as planets, the masses would actually be best constrained by RVs and not TTVs. This was disappointing because the promise of TTVs was to measure masses without the expensive RVs, but the limited results from the short amount of available data could not be helped.  

A discussion ensued about the possibility of another planet beyond the two visible planets. A planet near the external 2:1 resonance with KOI-377.02 (e.g., at 75 days) would be the most likely to make a difference in the TTVs, but it couldn't be a big difference within the constraints of the available data. We decided to postpone that analysis, particularly since we have a very good fit to the existing data.\footnote{To date, the Kepler-9 TTVs are fit extremely well by a model with only 2 planets, but essentially no 3-planet models have been investigated (besides the inclusion of Kepler-9d).} 

Using some best-fit models and extending them to $\sim$4 years, DF showed that the longer baseline would be able to measure mass ratios and orbital properties to within $\sim$1\%; this predication was validated by later analyses (e.g., \citep{2018A&A...618A..41F}). 

On May 27, Eric Ford sent out a detailed paper outline with assignments for all the sections. The working title is: "Kepler-N: A Sun-like star with multiple transiting planets and masses constrained by transit time variations."

On May 28, we discussed the issue of mutual inclinations. DF finds that even planets with mutual inclinations of up to 135 degrees can still fit the TTV data (as expected based on earlier results from \citep{2010ApJ...712L..86P}). We also get final spectroscopic properties of KOI-377 that showed that it is Sun-like in basically all respects. 

On May 29, we had a discussion on transit timing uncertainties and how they should be propagated. Based on a bootstrap analysis, the uncertainties are $\sim$80 seconds, but there was some preference to inflating the uncertainties by a global multiplier so that the reduced chi square would be $\sim$1. 

On May 31, Bill Welsh reported on his fully independent light-curve analysis that confirmed the TTV properties. It also measured depths and durations for each of the planets over time and these were consistent with constant. (The full four years of Kepler data do show subtle duration variations, discussed below.) 

By June 1, some of the coauthors had written up their subsection drafts. On June 2, we decided that the photodynamical model was just a proof of concept, but too slow to be used for detailed fitting; it was good for making a figure (eventually Figure 5 in H+10). On June 3, we held our regular KTMWG telecon including discussion on KOI-377, where everyone involved was asked to submit draft text for their sections. We discussed which journal to submit the results to. 
On June 4, Geoff Marcy sent 4 RV measurements to the group and quick analyses ensue. Some concern about missing intermediate planets was raised. Over the course of the next week, these observations changed due to different reductions of the data. These are some of the first high-precision RV measurements around such a faint star and even the final numbers were not consistent with additional later analyses, for a variety of possible reasons, see \citep{2019MNRAS.484.3233B}. 

On June 7, we discussed what to do with KOI-377.03. This early in the Kepler mission it would have taken a lot of work and checking to ensure that it was a real planet in the same system; we eventually decided to leave it as a candidate for this first paper. 

From June 8-10, most Kepler science team members were at the science team meeting in Aarhus, Denmark. We continued to assemble draft sections and figures. Willie Torres and Francois Fressin used their BLENDER model to confirm that KOI-377.01 and KOI-377.02 were not false positives. 

A search was done in the TrES dataset, which observed KOI-377, but no believable transit events were found.\footnote{Now that Kepler-9's TTVs are so well understood, it may be worth checking again for precovery data. However, as seen by the analysis of \citep{2018A&A...618A..41F}, the \emph{Kepler} data are so precise that ground-based observations even several years away do not provide significant improvement in planetary parameters.} 

Progress then slowed, partly because there were so many interesting Kepler systems to work on. 

At the June 18 KTMWG telecon, the group was still split on submitting to \emph{Science} or the \emph{Astrophysical Journal} and on when the paper should be submitted. MH, who was pretty clearly leading the paper at this point, proposed \emph{Science} with the reasoning that many of the details can be included in the Supplementary data and that this also might open up room for others to do independent follow-up papers.\footnote{The planned papers never really materialized as focus shifted to other interesting \emph{Kepler} systems and analyses.} Furthermore, this would be a nice follow-up to the \emph{Science} paper \citet{2005Sci...307.1288H} about the promise of TTVs. Finally, submitting to \emph{Science} was ``splashier'' and the KOI-377 result was at this point seen by the Kepler Science Team and Mission as a great way to generate good press. 

At the June 23 KTMWG telecon, with permission from the Kepler Science Council, MH (and the group) decided to submit the KOI-377 to \emph{Science}. The goal was to submit very soon so that it would be accepted by October 1st, allowing it to go into a NASA press conference shortly thereafter. Work began to identify how to compress the main text and what could go into separate papers. 

Emails on June 24 pointed out some concerns about the RV data and so plans are made to obtain 2 more observations. RVs were valuable to the paper because they are the primary constraint on the total mass (but not the planet-planet mass ratio) and because they would assuage concerns by readers who were not familiar with TTVs. 

On June 24, Jason Rowe sent the new Quarter 4 TTVs for KOI-377 that showed a continuation of the trends we observed so far. These data were added to the analysis and were those published in H+10. 

In early July, work continued on the details of the KOI-377 paper. NASA expressed interest in moving up the press event timeline to August, which would require some speedy work. Geoff Marcy provided 2 more RV measurements and DF folded them into his model. They shifted the masses somewhat, but otherwise everything looked great. Shortly thereafter, Lars Buchhave identified moonlight as the probable cause of the unusual RV bisectors, removing one of the last concerns to the KOI-377 paper. DF also presented some new results: GYr stability for the system and large resonance libration amplitudes.

On July 11, MH sent out a solid first draft in \emph{Science} format. Over the next couple of weeks, MH and coauthors put in a lot of editing work to get the paper in shape. MH, DF, and DR led the paper with many important coauthors from the KTMWG and the Kepler Science Team in general. At this time, it was common to include many members of the Kepler Team, even if their contributions to the specific scientific analysis were minimal; their significant preparatory work enabled the exciting science.  Many co-authors contributed additional analyses on a variety of aspects of the system; most of these were placed in the Supplementary Online Material (which ended up being 27 pages in length). 

The Kepler Science Council gave KOI-377.01 and KOI-377.02 the names Kepler-9b and Kepler-9c. KOI-377.03 would remain a candidate that will be confirmed via BLENDER in a near-term follow-up paper (see below). 

On July 27, only 11 weeks after first learning about KOI-377, MH submitted the KOI-377/Kepler-9 paper to \emph{Science}! The final name for the title was "Kepler-9: A System of Multiple Planets Transiting a Sun-Like Star, Confirmed by Timing Variations." 

On August 9, the first round of comments from three referees came back from the journal. In general, the referee's comments were very positive and recommended only minor changes. One referee rightly pointed out that we could not clearly infer resonance libration and that we should refer to the 2.023 period ratio as a near-commensurability.\footnote{Further work on \emph{Kepler} MTSs has shown that actual resonance occupation is, indeed, more complicated and that it is difficult to establish with TTVs (e.g., \citep{2016Natur.533..509M,2016AJ....152..105M}).} 

After rapid work to edit the paper and write a response to the referees, it was resubmitted to \emph{Science} on August 13. An additional very minor referee response was sent on August 18 to which we easily responded. The paper was officially accepted on August 19. The paper was published in Science Express on August 26 so that MH, DF, and DR could present Kepler-9 at the ``Detection and dynamics of transiting exoplanets'' conference at Observatoire de Haute-Provence in late August. \emph{Science} asks for Kepler-9 art to use on the cover and quick work allowed the October 1, 2010 \emph{Science} magazine to have a beautiful cover with two planets and a Sun-like star. The issue also included a nice ``Perspective'' on the Kepler-9 result from Greg Laughlin\citep{2010Sci...330...47L}.

\begin{figure}						 
    \begin{center}
    \includegraphics[trim={2in 6in 1in 1in},clip ]{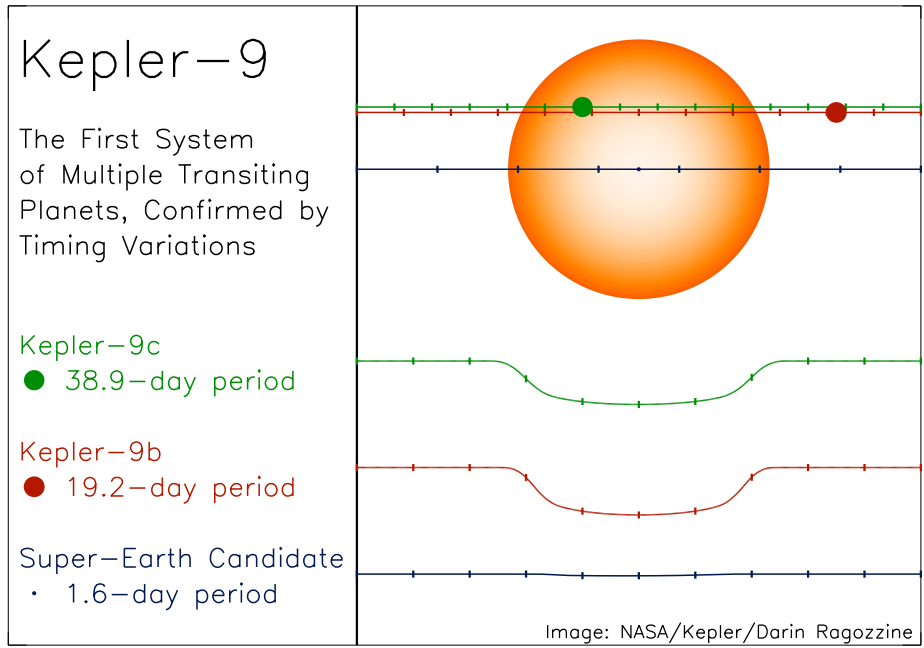}
	\caption{Kepler-9 was the first star with multiple confirmed transiting planets and the first star for which the planets showed clear TTVs. Kepler-9b in its 19.2-day orbit executes almost exactly two orbits during one 38.9-day orbit of Kepler-9c and this relationship amplifies the interaction between the planets, which is precisely measured by the Kepler Space Telescope. Another candidate planet, known as KOI-377.03 was not fully confirmed at the time of this press release, but was validated by BLENDER shortly thereafter \citep{2011ApJ...727...24T}. The upper right figure shows an example of how the Kepler-9 system might look if we could zoom in to the region around the star; the three planets cross the star in different ways, which are shown here to scale, along with the size of the planets (including tiny KOI-377.03 at the center of the star). Lightcurves for the three planets are shown in the lower right. Even though the depth of the transit for KOI-377.03 (about 240 parts per million) is barely the thickness of the line, it was well-detected by the precise Kepler telescope, illustrating its ability to find tiny Earth-size objects. Tick marks every hour show how the more distant planets (with longer periods) cross the star more slowly, in accordance with the celestial laws of motion discovered by Johannes Kepler, after whom the space telescope is named. A commenter on the Universe Today article ("polarisdotca") correctly points out that the hour-ticks are not correct on the transit light curves of Kepler-9b and 9c. The NASA press kit included a modified figure (now on the Wikipedia Kepler-9 page) that removed the credit statement from the image, though DR was still credited on the NASA website (\texttt{https://www.nasa.gov/mission\_pages/kepler/multimedia/images/kepler9.html}).}
	\label{DRKep9}
    \end{center}
\end{figure}

Associated with the August 26 \emph{Science Express} publication was a NASA press release. DR prepared a graphic of the system to go with the press release based on his photodynamical model. This is shown in Figure \ref{DRKep9}. While \emph{Science} does not allow statements of priority or novelty, Kepler-9 was the first publication of a star with more than one transiting planet and the first (statistically significant) example of TTVs. 

As TTVs were a new way of understanding exoplanets (for the public), NASA produced an animation\footnote{Now at \texttt{https://www.youtube.com/watch?v=-wG0Omed8Ow}.} that demonstrated how TTVs worked by comparing the periodicity of a single-planet system with deviations caused by a second planet. The Kepler-9 press release was well received and brought good press to the mission, exoplanets, and the authors. Using Wikipedia's Pageviews and Langviews analysis, we see that the Wikipedia pages on Kepler-9, Kepler-9b, or Kepler-9c have been accessed nearly 50,000 times as of February 2019.

\section{Kepler-9 since Holman et al. 2010}
H+10 was an auspicious launch of the study of MTSs and TTVs and there are now dozens of papers with thousands of citations on each of these two topics. Here we review major studies of the Kepler-9 system in particular. 

The same night as the Holman et al. 2010 \emph{Science Express} article, the BLENDER team lead by Willie Torres and Francois Fressin posted their paper on arXiv that ``validated'' KOI-377.03 as Kepler-9d\citep{2011ApJ...727...24T}.\footnote{''Validation'' refers to the fact that the candidate transit signal was identified as a \emph{bona fide} exoplanet by showing that the signal was far more likely to be from a small exoplanet than from an astrophysical small positive (as opposed to a ``confirmation'' which shows directly that the object has planetary mass).} At the time, Kepler-9d was the smallest exoplanet (though not necessarily the least massive). \citet{2011ApJ...727...24T} is submitted a week later and accepted a couple of months later. 

As discussed in \citep{RH10}, MTSs are particularly valuable for comparative exoplanetology because mass and radius ratios are much better measured than masses and radii and also because assumptions about planetary properties (e.g., bulk composition) are more likely to be correct for planets in the same system. An example of this is shown explicitly for the Kepler-9 system by \citet{2011A&A...531A...3H} who find that they can constrain the ratio of heavy element fractions better than each individual interior structure. 

Despite the continual downpouring of new \emph{Kepler} data that showed significant evolution in the Kepler-9 TTVs, it took 4 years before a new detailed TTV analysis was published by \citet{2014arXiv1403.1372D} followed by \citep{2014A&A...571A..38B}. (Some intermediate analyses were presented at conferences.) Both analyses found that, with a much more extensive TTV dataset, the planetary masses can be determined accurately and independently of RV data. Both\footnote{We note that \citep{2017AJ....154....5H} also retrieved similar masses.} also find that, when excluding RV data, the planets are only 55-60\% as massive as reported in H+10 implying very low densities of $\sim$0.2 g cm$^{-3}$.\footnote{When analyzing only the data available to H+10, they reach conclusions consistent with the original analysis. In addition, as predicted by H+10, the mass ratio of the planets is very tightly constrained to 0.6875 $\pm$ 0.0003 and most orbital properties are measured to $\sim$1\% precision.)} While some have construed this as a possible point of tension between RV and TTV mass estimates, this is more likely to be due to systematic errors in the RVs; Kepler-9 RV measurements were some of the very first precision measurements done on such a faint target. Recently, \cite{2019MNRAS.484.3233B} published 30 new RV observations from HARPS-N showing that the TTV and RV solutions are fully self-consistent as long as the H+10 RVs are removed from the analysis. This provides closure on the issue of differences in the TTV and the original Keck RV data by demonstrating that the Keck RV data must be affected by some unrecognized systematic errors. 

\citet{2018AJ....155...73W} obtained a new ground-based transit in September 2016, over three years since the final \emph{Kepler} data. They find a discrepancy of 45-minutes which is larger than expected, but not enough to suggest a serious violation of the existing models.

\citet{2018AJ....155...70W} obtained a Rossiter-McLaughlin measurement of the (projected) alignment between the stellar spin axis and the orbit of Kepler-9b to find it is well-aligned (though with moderate uncertainties). Kepler-9 thus joins the small list of MTSs with alignment measurements, all of which point to coplanarity between the stellar equator and planetary orbits. Such a result is not surprising and supports the hypothesis that protoplanetary disk interactions are strongly involved in the formation and evolution of these planets. 

At the present time, the best planetary parameters are given by \citet{2018A&A...618A..41F}, which exceeds previous analyses by 1) using a photodynamical model; 2) using \emph{Kepler}'s 1-minute ``Short Cadence'' data; 3) including several ground-based transits (9 of Kepler-9b and 4 of Kepler-9c); and 4) using GAIA's DR2 parallax measurement to help constrain the properties of the star. The photodynamical model is able to pick up on the slight Transit Duration Variations (though these were already clear in the tables of \citep{2016ApJS..225....9H}). In conjunction with a long-period TTV signal, the TDVs (automatically incorporated into the photodynamical model) allow  \citet{2018A&A...618A..41F} to detect the slow change in apparent inclination that will lead to Kepler-9c becoming non-transiting in the 2040-2060 time frame. We note that the inclusion of ground-based transits did not provide a significant improvement of the model over \emph{Kepler} Short Cadence data alone, even though these transits doubled the observational baseline. The \emph{Kepler} data are so extensive and precise that it will take several years before TTVs from the ground or other spacecraft provide significant improvement to the model parameters. This is unsurprising for a system that was chosen because of its high information content even with the earliest \emph{Kepler} data. 

The results of \citet{2018A&A...618A..41F} show that Saturn-sized Kepler-9b and Kepler-9c have some of the most precise densities known for any exoplanet at 0.439 $\pm$ 0.023 and 0.322 $\pm$ 0.017 g~cm$^{-3}$, respectively. These are significantly higher than \citet{2014arXiv1403.1372D} because of a smaller stellar radius and different inferred planetary radius ratios. Practically all orbital elements are measured to $\sim$1\% precision (as predicted in H+10), with the dominant uncertainties from the unknown stellar properties. 

Using instead the very precisely inferred mass ratio (0.68849 $\pm$ 0.0002) and estimating the radius ratio to be 0.97886 $\pm$ 0.0003, we infer a density ratio between the two planets of 0.7341 $\pm$ 0.001, i.e., measured to an incredible 0.14\%. Using this precise density ratio to learn more about the interiors at this level of precision would require considering other small effects like the planetary rotational and tidal bulges\citep{2009ApJ...698.1778R}.

\section{Conclusions and Further Questions}

Kepler-9 is exciting for both historical and scientific reasons. As the first multi-transiting system and first system to show TTVs, it established how \emph{Kepler} was going to revolutionize the study of precision exoplanetary dynamics. The strong planetary interactions and \emph{Kepler}'s incredible lightcurves allow for exquisite percent-level characterization of the orbital and physical properties of Kepler-9b and Kepler-9c. 

Having insight to the historical origin of the analysis of this system and reflecting on recent analyses, we conclude by imagining what might be possible for the future of Kepler-9 by listing outstanding questions that could deserve further investigation. 
\begin{itemize}
    \item Can all three planets of Kepler-9 be found in a single search for quasiperiodic signals? How many currently unknown planets (and systems!) would be turned up by such a search over all the Kepler data? 
    \item With the recently improved masses, radii, and densities (and the very precise ratios of these quantities), what can we infer about the detailed internal structure of Kepler-9b and 9c? How does that affect our interpretation of Kepler-9d? 
    \item What are the detailed upper-limits on additional planets throughout the Kepler-9 system? Is the two-planet-only TTV model fully justified? The new RV data from \citep{2019MNRAS.484.3233B} span 150 days and show no evidence for additional planetary signals, but upper limits are not quantitatively derived. 
    \item A small inner planet separated from two much larger planets is seen in Kepler-9, Kepler-18, Kepler-27, Kepler-297, GJ 876, and other systems. Is this a common system archetype and if so, why? How is it related to similar systems like the Kepler-31 system with a small inner planet and three larger exterior planets? 
    \item Are the properties of Kepler-9b and 9c consistent with core accretion? With their unusually large envelopes and low densities, they do not match the standard story of rapid accretion of an envelope onto a $\sim$10 Earth mass core. 
    \item The Kepler-9 planets are among the largest MTSs in this period range. While it is expected that the first planets analyzed would be the largest (and thus highest SNR) planets, Kepler-9 still could provide insight on the upper end of the radius distribution of MTSs. 
    \item How did the Kepler-9 system form? Is the hypothesis of formation and then significant migration consistent with all the observations? 
\end{itemize}

We thank the \emph{Kepler} Mission and Science Team for the incredible journey that we've been privileged to participate in. We thank the members of the Kepler TTV/Multis Working Group for many years of productive and enjoyable discussions. We thank the co-authors of the Kepler-9 paper for their valuable contributions. We thank an anonymous referee, Eric Ford, Dan Fabrycky, Jack Lissuaer, and Joann Eisberg for suggestions that improved the quality of this work.



\bibliographystyle{model1a-num-names}

\bibliography{all}

\end{document}